**Dualities and Twins: Reflections on Hapgood**

Adrian Kent

Centre for Quantum Information and Foundations, DAMTP, University of Cambridge, U.K.

Perimeter Institute for Theoretical of Physics, 31 Caroline Street N, Waterloo, Ontario, Canada

**Abstract:** These programme notes were written for a production of "Hapgood" at the Hampstead Theatre, London in December 2015.   I thank Tom Stoppard for helpful suggestions.

KERNER: *Every time we don't look, we get wave pattern. Every time we look to see how we get wave pattern we get particle pattern.*

BLAIR: *How?*

KERNER: *Nobody knows. Somehow light is continuous and also discontinuous. The experimenter makes the choice. You get what you interrogate for. And you want to know if I'm a wave or a particle.*  [1]

Humans have a natural scientific instinct.   Evolution has taught and equipped us to try to make sense of the social and physical worlds we live in through developing, testing and refining models.   If you jump off your roof, I think you may well hit the ground quite hard: let's see what happens with this brick.   And so on.     In developing this natural inclination to explore and understand the world into theories of physics, we've learned that the formal language of mathematics gives us models that work astonishingly well.   This "unreasonable effectiveness of mathematics in the natural sciences", as the Hungarian-American physicist Eugene Wigner put it [2], leads some to think that our universe is not just well described by mathematics, but is – in some sense that is hard to properly articulate – a fundamentally mathematical object.   We've also learned that the intuitions we have developed from physics at human scales and larger – the physics of bullets, cricket balls and planets, or water and sound waves – just don't apply to objects at very small scales.   To describe atoms, electrons, and tiny pulses of light needs a new theory, quantum theory, with its own mathematical laws and its own conceptual framework.

The famous experiment [3,4] Kerner describes illustrates this beautifully.    To decide what happens if you shine light through two slits in a wall, you need some sort of mental model.    Here's one: maybe light is actually made up of tiny elementary pulses of light – corpuscles, as Isaac Newton and his contemporaries called them.   And maybe they follow straight paths, like the bullets to which Kerner compares them earlier in the scene.   If so, you'd expect to see light coming through the two slits, and nowhere else.     But that's not what happens [4]: you also see dark and light fringes in the areas between and either side of the slits.     Well, you can see something like that when water waves hit a breakwater with two gaps.    So we could instead try modelling light as something like a water wave.    But that doesn't work either: when we look closely enough we see that the light hits our detectors in discrete pulses at definite points, very much like Kerner's bullets and very unlike a continuous water wave that arrives across a wide area of the beach.     Popularisers sometimes say that light is both a particle and a wave, but it's more accurate to say that it's neither.

So, why not look more closely and work out what's really going on?    After all, you would think, *something* must be going through at least one of the slits.    If we put detectors just behind each slit we can see what's going through them, and if even that doesn't fully unravel the mystery, we can put detectors everywhere, and piece together what's really happening. The trouble, as Kerner says, is that the act of looking changes the experiment *and changes the results*. We don't see the dark and light fringes any longer, so whatever we learn in this new experiment doesn't help to explain how they arose in the previous one. This is a quite general feature of quantum theory. The answers depend on the questions, and observing a physical system changes its future behaviour: in fact, by choosing the right sequence of observations, we can make it follow any path we wish [5].

Another striking aspect of quantum theory is the fundamental role played by dualities of various kinds. We can describe an object in terms of its position or momentum, but can never specify both. Electric and magnetic fields, which seem at first sight quite different, turn out to be dual aspects of a unified field theory. The phenomenon of quantum entanglement, first studied by the Austrian physicist Erwin Schrodinger in 1935 [6], means that separated pairs of quantum particles behave in a way that, according to our (once again, let me stress, fallible!) pre-quantum mental models, requires a sort of sub-atomic telepathic communication: in Einstein's phrase, "a spooky action at a distance". The anti-particles with which Kerner works are mirror images of their counterparts, with the same mass and opposite electric charge. More speculatively, but intriguingly, recent theories suggest that there may be a duality between gravity and the other fundamental forces in quantum theory [7].

Many of these features of quantum theory have pleasing parallels in our individual and collective psychologies. The act of observing a situation can certainly change it. Asking a question can maybe not only evoke an answer but create a new psychological or sociological truth. Human nature too has its dualities, of course, and it is not only in quantum theory that we are sometimes tempted to see something uncanny about twins and doppelgangers. Indeed, Wolfgang Pauli, one of founders of quantum theory, had a long collaboration with Carl Jung [8], and both thought they saw deep connections between their fields. It's a very big claim: after all, quantum systems follow precise mathematical laws; humans, not so much. Moreover, we don't yet entirely understand what quantum theory says even about the material world. You'll find lively debates at most quantum foundations conferences. Proponents of a "many worlds interpretation" think quantum theory implies there are real parallel universes corresponding to everything that could possibly have happened or possibly will [9]; others argue there really are elementary particles bobbing along guided by quantum waves [10,11]; subjectivists reject any standard picture of physical reality and suggest the theory is all about our experiences [12]; collapse model theorists propose a new post-quantum theory to unify small and large scale physics [13]. Given all this, I'd rather not claim too much. Our individual and collective consciousnesses seem deeply mysterious to me. Maybe they can't be fully understood scientifically. But we try to make sense of them by metaphors and analogies, in novels, poems, films and plays -- and when quantum theory plays an entertaining part, I relish it.